\documentclass[aip,reprint,onecolumn]{revtex4-1}

\usepackage{epstopdf}
\usepackage{hyperref}
\usepackage{url}
\usepackage{amsmath}
\usepackage{amssymb}
\usepackage{graphicx}
\usepackage{subcaption}
\usepackage{tabularx}
\usepackage{makecell}
\usepackage{booktabs}
\usepackage{amsthm}
\usepackage{listings}
\usepackage{paralist}
\usepackage{color}
\usepackage{siunitx}
\usepackage{multirow}

\DeclareMathOperator*{\argmin}{arg\,min}
\DeclareSIUnit{\cal}{cal}
\DeclareSIUnit{\kcal}{\kilo\cal}

\draft 

\begin{document}

\title{Computing Committor Functions for the Study of Rare Events Using Deep Learning} 

\author{Qianxiao Li}
 \email{matlq@nus.edu.sg}
\affiliation{Department of Mathematics, National University of Singapore, Singapore 119076}
\affiliation{Institute of High Performance Computing, A*STAR, Singapore 138632}
\author{Bo Lin}
 \email{E0046836@u.nus.edu}
\affiliation{Department of Mathematics, National University of Singapore, Singapore 119076}
\author{Weiqing Ren}
 \email{matrw@nus.edu.sg}
\affiliation{Department of Mathematics, National University of Singapore, Singapore 119076}


\begin{abstract}
The committor function is a central object of study in understanding transitions between metastable states in complex systems. However, computing
the committor function for realistic systems at low temperatures is a challenging task, due to the curse of dimensionality and the scarcity
of transition data. In this paper, we introduce a computational approach
that overcomes these issues and achieves good performance on complex benchmark problems with rough energy landscapes.
The new approach combines deep learning, data sampling
and feature engineering techniques. This establishes an alternative practical method for studying rare transition events between 
metastable states in complex, high dimensional systems.
\end{abstract}

\pacs{}

\maketitle 

\section{Introduction}
\label{introduction}

Understanding transition events between metastable states is of great importance 
in the applied sciences.
Well-known examples of the transition events include
nucleation events during phase transitions, conformational changes of bio-molecules, dislocation dynamics in crystalline solids, etc.
The long time scale associated with these events is a consequence
of the disparity between the effective thermal energy and typical energy barrier of the systems. The dynamics proceeds by long waiting periods around metastable states followed by sudden jumps from one state to another. For this reason, the transition event is called rare event.
%
The main objective in the study of rare events is to understand
the transition mechanism, such as the transition pathway and transition states.
Some numerical methods have been proposed for this purpose, among which the well-known ones include the nudged elastic band method, \cite{jonsson1998nudged} the string method, \cite{weinan2002string,weinan2007string,ren2005transition} the action-based method, \cite{olender1996calculation}
and the transition path sampling technique, \cite{bolhuis2002transition,dellago2002transition}
accelerated molecular dynamics, \cite{voter1997hyperdynamics} etc.

One object that plays an important role in understanding the transition event 
is the committor function. This is a function which is defined
in the configuration or phase space
and describes the progress of the transition.
Most of the interesting information regarding the transition can be extracted 
from the committor function. \cite{weinan2005cpl,ren2005transition,weinan2010ARPC}
For example, the transition states lie in the region where the committor value 
is around $1/2$; the transition path ensemble and the transition rate can be computed
as well, based on the committor function and the equilibrium probability distribution
of the system. Thus, developing efficient numerical methods to compute the committor
function is an important problem for understanding rare events.

The committor function has a very simple mathematical description - it satisfies the backward Kolmogorov equation. However, it is very difficult to compute in practice, due to the curse of dimensionality. For example, the equation needs to be solved 
in the configuration space of dimension $3N$ for a molecule consisting of $N$ atoms. 
Traditional numerical approaches, such as finite difference or finite element methods, are computationally prohibited even when $N=2$ or 3,
and obviously out of the question for realistic physical systems.
Alternative methods have been proposed. In the transition path sampling
technique, \cite{bolhuis2002transition,dellago2002transition}
the committor function is computed using Monte Carlo methods.
In the finite-temperature string method, \cite{ren2005transition} the problem 
is reformulated into an equivalent variational form, 
which is minimized in a particular function space.
In the work of Ref.~\onlinecite{lai2018point}, the Kolmogorov equation for the committor function is solved using a point cloud discretization.

More recently, it is proposed to compute the committor function using neural networks. \cite{khoo2019solving} Satisfactory numerical results were obtained 
for a model problem in $10$ dimensions.
The success of this approach hinges on the availability of data, especially in the transition state region where the committor function changes sharply from 0 to 1. As the term ``rare event'' suggests,
such data is rarely available. In Ref.~\onlinecite{khoo2019solving},
the data was generated by solving the underlying Langevin dynamics at the physical temperature. While this works well when the physical temperature is high in which case transitions are easily observed,
it becomes less efficient when the temperature is low, i.e. the case of rare events.

In the current work, we propose to use importance sampling techniques 
to overcome this difficulty. We consider two methods. In the first one, 
we generate the data using the Langevin dynamics at an artificial temperature. 
This temperature is high enough so that the thermal energy becomes close or even
comparable to the energy barrier. In this case,
the transition event between the metastable states becomes less rare or even frequent.
This enables us to collect sufficient amount of data in the transition state region.
The difference between the physical temperature and the artificial temperature 
is accounted for by the likelihood ratio in the objective function to be minimized.
In the second method, we sample the data using metadynamics. 
The relevant potential wells on the energy surface are first filled
using localized Gaussian functions. This effectively lowers the energy barrier
between the metastable states. The data can then be sampled efficiently following 
the Langevin dynamics on the modified potential.

In addition to these sampling methods, in this work we also introduce 
new methods to impose the boundary conditions for the committor function, 
and employ collective variables as the input features for the neural network.

The paper is organized as follows. The background and problem formulation are presented in Section~\ref{background}. In Section~\ref{method}, we introduce the key ingredients of our method, including the method of imposing the boundary conditions, data sampling techniques, and the neural network architecture.
Numerical results for two examples are presented in Section~\ref{experiments}.
Finally, we draw conclusions in Section~\ref{conclusion}.

\section{Background}
\label{background}

The typical starting point of transition modeling is the specification of a potential energy function $V:\Omega \subset \mathbb{R}^n \rightarrow \mathbb{R}$, which takes as inputs the microscopic configuration $x$ of the system (e.g.\,the positions of the constituent atoms of a bio-molecule). The subset $\Omega$ is the configuration space under consideration. Through standard statistical
mechanics arguments, the probability density of the system's configuration is determined by the potential energy function $V$ via the Boltzmann-Gibbs distribution
\begin{equation}\label{equi}
p(x) = \frac{1}{\mathcal{Z}} e^{-\beta V(x)},
\end{equation}
where $\beta=1/k_{B}T$ is the inverse temperature and $\mathcal{Z}=\int_{\Omega}e^{-\beta V(x)}dx$ is the normalization factor, or partition function.

To study dynamical properties, one may consider the noisy gradient flow 
induced by $V$ in the form of the  over-damped Langevin equation:
\begin{equation}\label{lag}
\dot{x}(t)=-\nabla V(x(t))+\sqrt{2\beta^{-1}}\eta(t),
\end{equation}
where $\eta$ is a white noise. One can check that Eq. \eqref{lag} 
has the invariant distribution in Eq. \eqref{equi}. Through this dynamical model, 
it is clear how one can then introduce the notion of stability of the system: 
local minima, or a collection of local minima, of $V$ correspond
to metastable configurations; at low temperatures such that the thermal energy 
is much lower than the energy barrier, 
the system will remain in these configurations for exponentially long times before
making transition to another such configuration. Our principal goal is to study the transition dynamics between two metastable configurations.

For two given distinct metastable regions $A, B \subset \Omega$, $A\cap B = \varnothing$,
consider the first hitting times 
\begin{equation}\label{hittingtimes}
\begin{aligned}
\tau_A(x)&= \inf \{ t \geq 0 : x(t) \in A , x(0) = x \}, \\
\tau_B(x)&= \inf \{ t \geq 0 : x(t) \in B , x(0) = x \}.
\end{aligned}
\end{equation}
The committor function $q : \Omega \rightarrow [0,1]$ is defined by
\begin{equation} \label{qdef}
q(x) = \text{Prob}\{ \tau_B(x) < \tau_A(x) \},
\end{equation}
i.e. $q(x)$ is the probability that the system initiated at $x$ first reaches $B$ rather than $A$.  
It can be shown that much of the vital information regarding transition pathways and rates can be extracted from the committor function. \cite{weinan2005cpl,weinan2010ARPC} 

Obviously, $q$ takes the value 0 in $A$ and 1 in $B$. In the domain 
$\Omega\setminus (A\cup B)$, $q$ satisfies the backward Kolmogorov equation
with Dirichlet boundary conditions, \cite{gardinerhandbook}
\begin{equation}\label{bK}
\begin{cases}
\nabla V\cdot\nabla q- {\beta}^{-1} \Delta q = 0,\quad x\in\Omega\setminus (A\cup B),\\
q(x)=0,\ x\in \partial A;\quad q(x)=1,\ x\in\partial B,
\end{cases}
\end{equation}
where $\nabla=(\frac{\partial}{\partial x_{1}},...,\frac{\partial}{\partial x_{n}})$ is the gradient operator, $\Delta=\sum_{i=1}^n\frac{\partial^2}{\partial x_{i}^2}$ 
is the Laplace operator, $\partial A$ and $\partial B$ denote the boundary 
of $A$ and $B$, respectively.
In addition, we impose the boundary condition $\nabla q \cdot \mathbf{n}=0$ on $\partial \Omega$. Eq. \eqref{bK} has an equivalent variational formation. Specifically, 
the committor function is also the solution to the variational problem
\begin{equation}\label{VP}
\min_{q} \frac{1}{Z}\int_{\Omega\setminus (A\cup B)}
\lvert \nabla q(x)\rvert^2 e^{-\beta V(x)}dx,
\end{equation}
where $Z = \int_{\Omega \setminus (A\cup B)} e^{-\beta V(x)} dx$, 
subject to the boundary conditions
\begin{equation}\label{bdc}
q(x)=0,\ x\in \partial A;\quad q(x)=1,\ x\in\partial B.
\end{equation}

The main challenge in using Eq. \eqref{bK} to compute $q$ is the curse of dimensionality: as the dimension $n$ of the configuration $x$ increases, the computational complexity of classical finite difference and finite element methods increase exponentially, thereby prohibiting the use of these methods for realistic models. Although the variational formulation somewhat ameliorate this issue, at low temperatures (large $\beta$), which is the regime of interest for rare transition events, solving the minimization problem in \eqref{VP} 
becomes more challenging as the measure $Z^{-1} e^{-\beta V}$ becomes more ``singular'', 
i.e. the density is more concentrated near the minima of the potential energy.

In the next section, we introduce our proposed method of combining deep learning and data sampling to efficiently compute $q$ for high-dimensional systems and at low temperatures.

\section{Methods}
\label{method}

The numerical method is based on the variational formulation \eqref{VP}-\eqref{bdc}.
We first reduce this constrained minimization problem 
into an unconstrained problem by choosing a particular form for $q(x)$.
Then we use a deep neural network to parameterize the function to be minimized, 
thereby convert the minimization problem into an unsupervised learning problem.
This is followed by the discussion of the key issue of the method - 
the sampling of training data for the neural network.

%

\subsection{Imposing the Boundary Conditions}
\label{relax}

In the earlier work, \cite{khoo2019solving} the boundary conditions 
were imposed by minimizing a modified functional with an additional penalty term.
While this is straightforward to implement, the introduction of an additional 
penalty parameter may require a careful tuning. Moreover, it requires sampling 
additional data on the boundaries. We proceed differently by introducing 
a particular form of $q$ that naturally satisfies 
the boundary conditions~\eqref{bdc}. Specifically, we consider the composite form
\begin{equation}\label{form}
q(x)=\left(1-\chi_{A}(x)\right)
\left[(1-\chi_{B}(x))\tilde{q}(x)+\chi_{B}(x)\right],
\quad x\in \Omega\setminus (A\cup B),
\end{equation}
where $\chi_{A}$ and $\chi_{B}$ are smooth functions such that 
$\chi_{A}(x)|_{\partial A}=1$, $\chi_{A}(x)|_{\partial B}=0$ 
and $\chi_{B}(x)|_{\partial B}=1$. One can easily verify that
$q(x)$ in Eq. \eqref{form} satisfies the boundary conditions 
$q(x)|_{\partial A}=0$ and $q(x)|_{\partial B}=1$.

A convenient choice for $\chi_{A}$ and $\chi_{B}$ is to use the mollified 
indicator functions
\begin{equation}\label{asum}
\chi_A(x) = \left\{
\begin{array}{ll}
1, & x\in \partial A, \\
0, & x \in \Omega\setminus A^{\epsilon},
\end{array} \right.
\quad
\chi_B(x) = \left\{
\begin{array}{ll}
1, & x\in \partial B, \\
0, & x \in \Omega\setminus B^{\epsilon},
\end{array} \right.
\end{equation}
where $A^\epsilon$ and $B^\epsilon$ are two sets expanded from $A$ and $B$, respectively
\begin{equation}
\begin{split}
A^{\epsilon} = \left\{ x\in \Omega:\ \inf_{y\in A}\lvert x-y\rvert\leq \epsilon\right\},
\quad	
B^{\epsilon} = \left\{ x\in \Omega:\ \inf_{y\in B}\lvert x-y\rvert\leq \epsilon\right\}.
\end{split}
\end{equation}
Away from $\partial A$ and $\partial B$, $\chi_A$ and $\chi_B$ changes smoothly from 1 to 0 in a region of width $\epsilon$, respectively. 
With this choice, $q$ agrees with $\tilde{q}$ outside $A^\epsilon$ 
and $B^\epsilon$,
\begin{equation}
q(x)=\tilde{q}(x), \quad\text{for } x\in\Omega\setminus (A^{\epsilon}\cup B^{\epsilon}).
\end{equation}
The constrained minimization problem \eqref{VP}-\eqref{bdc}
now reduces to an unconstrained problem
\begin{equation}\label{VP1}
\min_{\tilde{q}} \frac{1}{Z}\int_{\Omega\setminus (A\cup B)}
\lvert \nabla q(x)\rvert^2 e^{-\beta V(x)}dx,
\end{equation}
where $q(x)$ takes the form \eqref{form}.

\begin{figure}[t]
\centering
\includegraphics[width=.8\linewidth]{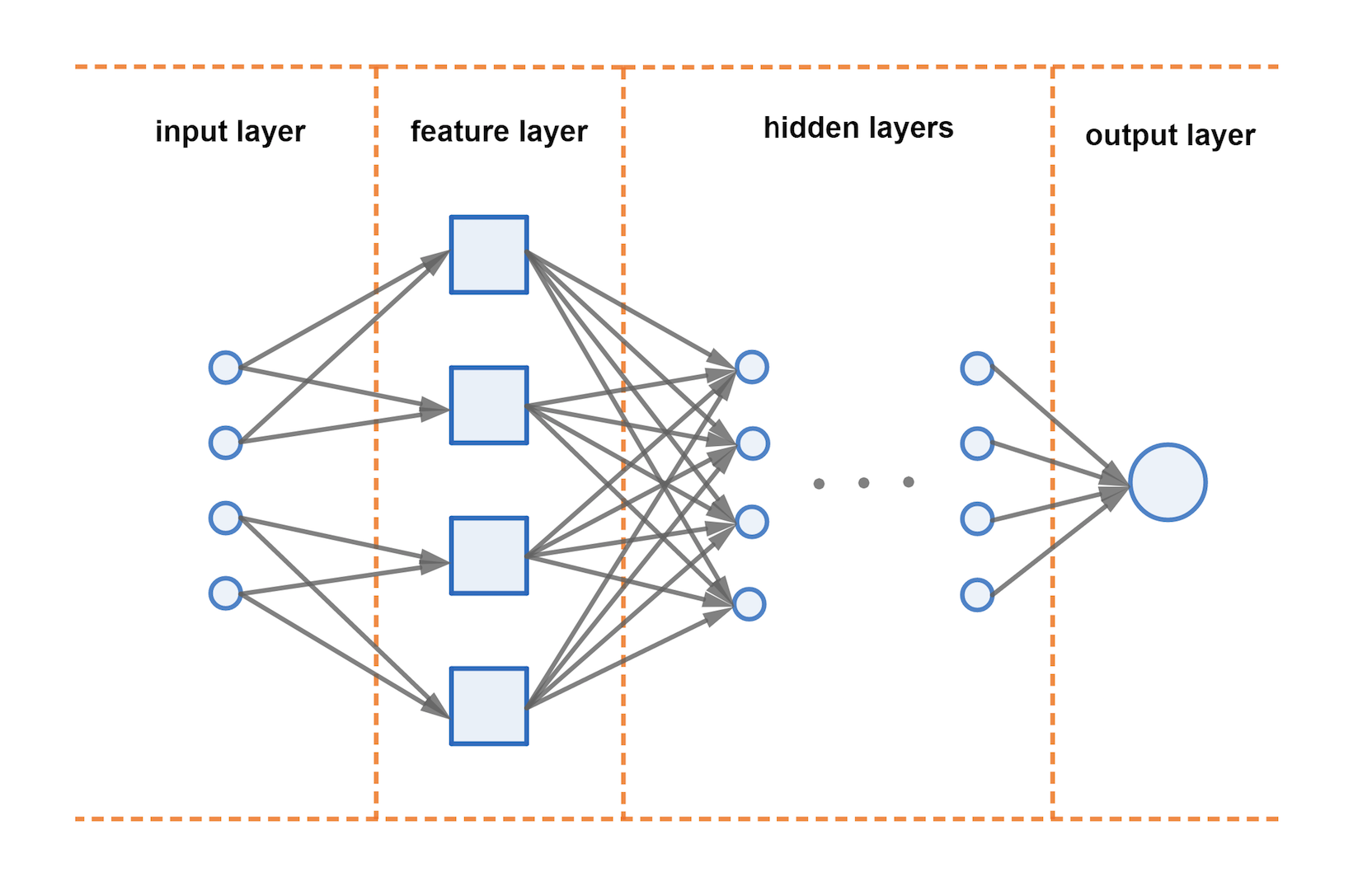}
\caption{Diagram of neural network structure with feature layer.
	The feature layer consists of collective variables information.
	The connection between the input layer and the feature layer is
	determined by the definition of the collective variables,
	whereas the feature layer and the first hidden layer are 
         fully connected. A sigmoid function is applied to the output layer.}
 \label{NN2}
\end{figure}

\subsection{Parameterization of the Committor Function}

We use a deep neural network to parameterize the function $\tilde{q}$, 
hence approximating the committor function $q(x)$ as
\begin{equation}\label{form2}
  q_{\theta}(x)=\left(1-\chi_{A}(x)\right)\left[(1-\chi_{ B}(x))
   \tilde{q}_{\theta}(x)+\chi_{B}(x)\right],\quad x\in \Omega\setminus (A\cup B),
\end{equation}
where $\{\theta\}$ are trainable variables of the neural network.
Specifically, given input vector $x\in \Omega\setminus (A\cup B)$, the neural network with $L$ hidden layers is defined as
\begin{equation}
\tilde{q} (x) = F_{L+1}\circ F_{L} \circ \cdots \circ F_{2} \circ F_{1}(x),
\end{equation}
where ``$\circ$'' denotes the composition of functions, $F_{i}(\xi)=\Phi_{i}(W_{i}\xi+b_{i})$ for $i=1,...,L+1$ and $\Phi_{i}$'s, $W_{i}$'s and $b_{i}$'s refer to the activation functions, weight matrices and bias vectors, respectively. 
Since the range of the committor function is [0, 1], we use a sigmoid activation for the output layer, i.e. $\Phi_{L+1}(\xi)=1/(1+\exp(-\xi))$. 

Very often, only a few coarse-grained variables, called the collective variables and denoted by $(z_{1}(x),z_{2}(x),...,z_{m}(x))$ with $m\ll n$, play a major role in the transition event. For example, in conformational changes of bio-molecules, it is often that only a few torsion or bond angles are sufficient to characterize the transitions between metastable states. In this situation, the committor function can be well approximated
by a function of these collective variables
\begin{equation}
q(x) \approx f(z_{1}(x),...,z_{m}(x)).
\end{equation}
This is a form of dimensional reduction or feature engineering,
and the choice of $\left\{z_i\right\}$ often allows one to build some degree 
of physical knowledge into the parameterization of $q$.
To that end, we shall make use of collective variables in the design of
the first input transformation layer of the neural network. 
The architecture of the network is summarized in Fig.~\ref{NN2}.

With the neural network parameterization, 
the minimization problem~\eqref{VP1} can be posed as an unsupervised learning problem
\begin{equation}\label{newVP}
\argmin_{\theta} \frac{1}{Z}\int_{\Omega\setminus (A\cup B)} 
\lvert \nabla_{x} q_{\theta}(x)\rvert^2 e^{-\beta V(x)}dx,
\end{equation}
where $q_\theta(x)$ takes the form~\eqref{form2}.

\subsection{Sampling Training Data for  the Neural Network}
\label{importance}

The objective function in Eq. \eqref{newVP} is in the form of an expectation
\begin{equation}
\frac{1}{Z}\int_{\Omega\setminus (A\cup B)} 
\lvert \nabla_{x} q_{\theta}(x)\rvert^2 e^{-\beta V(x)}dx 
= \text{E}_{X\sim p_\beta} \left[\lvert \nabla_{x} q_{\theta}(X)\rvert^2\right].
\end{equation}
A conventional approach to solve this type of learning problem is to use the stochastic gradient descent (SGD) algorithm, in which the expectation is approximated by sample average
\begin{equation} \label{eq:average}
\text{E}_{X\sim p_\beta} \left[\lvert \nabla_{x} q_{\theta}(X)\rvert^2\right] \approx 
\frac{1}{N}\sum_{k=1}^N \lvert \nabla_{x} q_{\theta}(X^{(k)})\rvert^2,
\end{equation}
where $X^{(k)}$'s are independent samples from the 
distribution $p_{\beta}(x) = Z^{-1} e^{-\beta V(x)}$.

The sampling of $p_{\beta}$ at low temperature $T$ (large $\beta$) poses a 
serious challenge. Naive sampling using the dynamics~\eqref{lag} 
at low temperatures yields very few transition events and hence the distribution 
$p_{\beta}$ is not explored well enough to solve~\eqref{newVP}. 
Specifically, with very few observed transition events, the majority of 
the sampled data will be clustered near the metastable states $A$ and $B$, 
with very few distributed in the region of our interest, 
i.e. the transition state region which lies in between $A$ and $B$.
This will lead to a poor estimate of the committor function.

The difficulty is caused by the disparity between the thermal energy $k_B T$ 
and the energy barrier between the metastable states. Next, we propose two 
sampling methods to overcome this difficulty. The first one is to use 
an artificially increased temperature, and the other one is based on
metadynamics which modifies the potential to lower the energy barrier.

\subsubsection{Sampling Data at Artificial Temperature}

We consider another form of the problem~\eqref{newVP}
\begin{equation}\label{newVP2}
\argmin_{\theta} \int_{\Omega\setminus (A\cup B)}
\lvert \nabla_{x} q_{\theta}(x)\rvert^2 e^{-(\beta-\beta')V(x)} p_{\beta'}(x)dx,
\end{equation}
where $\beta'=1/k_B T'$, with $T' > T$ and $p_{\beta'}(x) = {Z'}^{-1} e^{-\beta' V(x)}$ in which $Z'$ is the normalization factor.

Solving the new problem~\eqref{newVP2} requires sampling
$p_{\beta'}$, for example, 
by simulating the Langevin dynamics~\eqref{lag} at the high temperature $T'$.
This becomes much more efficient when $T'$ is sufficiently large,
as energy barriers are much easier to cross at higher temperatures.
In essence, this is a form of importance sampling, and the function
$e^{-(\beta-\beta')V(x)}$ is the likelihood ratio associated with the change of measure. Note that the choice of importance sampling measure $p_{\beta'}$ is not arbitrary. Its purpose is to produce enough sample points in the \textit{a priori} unknown transition region for us to estimate the committor function at low temperature $T$. Generic sampling measures (e.g. uniform) will not achieve this goal in moderately high dimensions.

\subsubsection{Sampling Data Using Metadynamics}

In this approach, an external potential $V_G$ is added to the potential $V$
to lower the energy barrier thereby facilitate transitions between the metastable states. 
Correspondingly, we consider the following equivalent form of the problem~\eqref{newVP}
\begin{equation}\label{metaVP}
 \argmin_{\theta} \int_{\Omega\setminus (A\cup B)} 
\lvert \nabla_{x} q_{\theta}(x)\rvert^2 e^{\beta V_{G}(x)}p_{G}(x)dx,
\end{equation}
where $p_{G}(x)=Z_{G}^{-1} e^{-\beta \left[V(x)+V_{G}(x)\right]}$ 
is the equilibrium distribution for the modified potential 
and $Z_{G}$ is the normalization factor.
We use metadynamics to construct the external potential $V_G$.

Metadynamics was developed to enhance the sampling in molecular dynamics simulations
and construct the free energy landscapes. \cite{laio2002escaping,barducci2011metadynamics}
The idea is to fill the potential wells containing the metastable states 
by depositing localized Gaussian functions following the dynamics. 
We denote the frequency of deposition by $\tau^{-1}$ and
use Gaussian functions in collective variables. Then 
at time $t$, the external potential that was added to $V$ is given by
\begin{equation}\label{meta_po}
V_{G, t}(x) = \sum_{t'=0,\tau,2\tau,...}^{t'< t} 
w\exp\left(-\sum_{i=1}^{d}\frac{(S_{i}(x)-S_{i}(x_{t'}))^2}{2\sigma_{i}^2}\right),
\end{equation}
where $w$ and $\sigma_{i}$'s are parameters that control the height and width of
the Gaussian respectively, $x_t$ denotes the trajectory on the time-dependent potential
$V+V_{G,t}$, and $\left\{S_i(x), i=1,\dots, d\right\}$ 
are some coarse-grained variables.
Since the system evolves towards local minima of the modified potential, 
the basins containing the metastable states are filled after sufficiently long time;
as a result, the transitions take place more frequently on the modified potential
compared to the original system. Thus metadynamics provides an efficient way to sample 
transition events even at low temperatures. 

In our simulation, we first fill the potential wells of $V$ using Eq. \eqref{meta_po}
up to a certain time $t^*$ when the system can easily hop over the potential barrier. 
This gives the external potential $V_G$ in Eq. \eqref{metaVP}: 
$V_G(x) = V_{G, t^*}(x)$. Afterwards, we sample 
data for the distribution $p_{G}(x)$
using the Langevin dynamics with the modified potential $V(x)+V_{G}(x)$. 
These data are used to solve the problem \eqref{metaVP}.

\section{Numerical Examples}
\label{experiments}

We demonstrate the effectiveness of the proposed numerical method using two benchmark problems: one is the Mueller potential extended to high dimensions, the other is the isomerization of alanine dipeptide.

\subsection{ Extended Mueller Potential}
We first consider the Mueller potential embedded in the $10$-dimensional ($10$D) space, \cite{khoo2019solving}
\begin{equation}\label{R10}
V(x)=V_{m}(x_{1},x_{2})+ \frac{1}{2\sigma^2}
\sum_{i=3}^{10} x_{i}^2, \quad x\in \mathbb{R}^{10}
\end{equation}
where  $V_m(x_1,x_2)$ is the rugged Mueller potential in two dimensions (2D),
\begin{eqnarray}\label{Mueller}
V_{m}(x_1,x_2) &=& \sum_{i=1}^4  D_{i}\exp[a_{i}(x_1-X_i)^2+b_{i}(x_1-X_i)(x_2- Y_i)+c_{i}(x_2-Y_i)^2 ]  \nonumber\\
&& +\gamma \sin(2k\pi x_1)\sin(2k\pi x_2),
\end{eqnarray}
and a harmonic potential is used in each of the other eight dimensions.
The parameters $\gamma$ and $k$ control the roughness of the energy landscape, and $\sigma$ controls the scale of the quadratic terms.
In this example, we use $\gamma=9,\ k=5,\ \sigma=0.05$.
The other parameters are taken from Ref.~\onlinecite{lai2018point}.

\begin{figure}[t!]
\centering
\includegraphics[width=.55\linewidth]{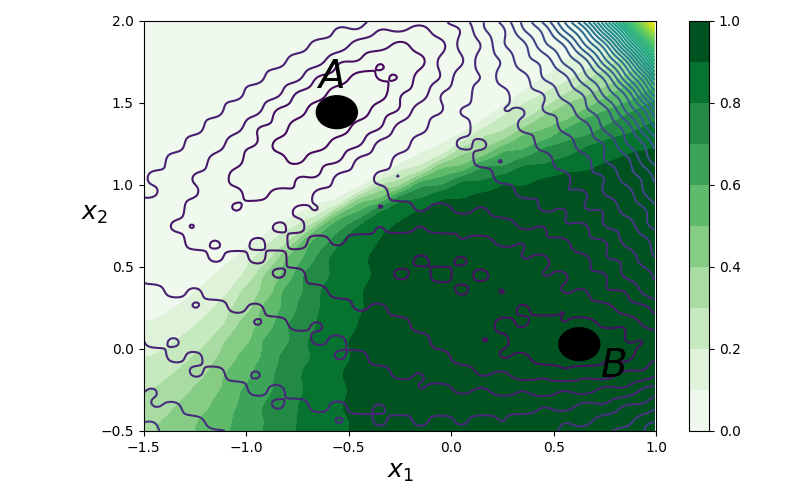} 
\includegraphics[width=.55\linewidth]{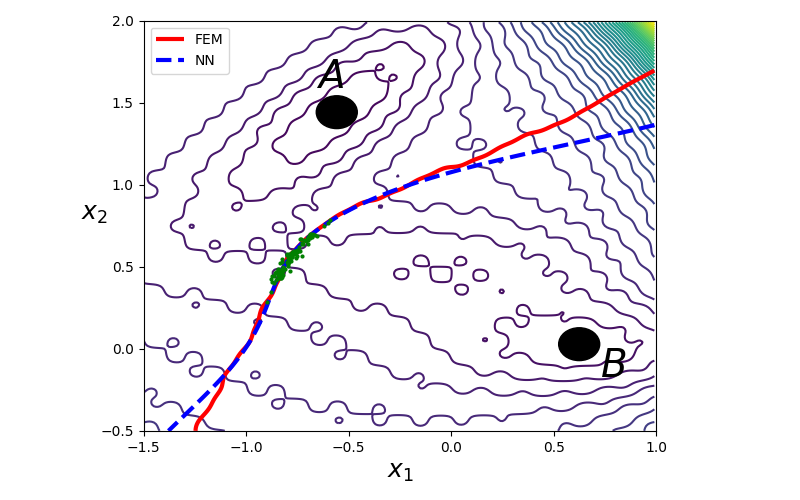}
\includegraphics[width=.55\linewidth]{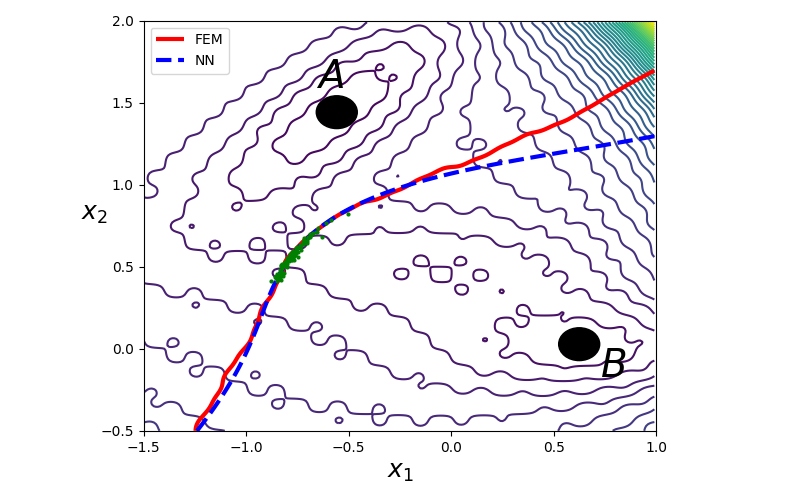}
\caption{
 {\it Upper panel:} Contour plots of rugged Mueller potential (solid lines) and
 the committor function $q_{m}$ computed using FEM.
 {\it Middle and lower panels:} The $1/2$-isosurface
 (projected onto the $(x_1, x_2)$ plane) of the neural network approximation
$q_\theta$ obtained using data sampled at higher temperature 
(middle panel, dashed line) and using data sampled from metadynamics 
(lower panel, dashed line), respectively. 
The neural network contains one hidden layer with $20$ nodes.
The numerical results are compared with the $1/2$-isosurface of $q_m$ 
(solid lines). The discrete points are the sampled $100$ transition states.
}
	\label{fig2}
\end{figure}

We first use the finite element method (FEM)
to solve the backward Kolmogorov equation~\eqref{bK} 
for the 2D rugged Mueller potential $V_m$ on $\tilde{\Omega}=[-1.5,1]\times[-0.5,2]$
by the solver FreeFem++. \cite{MR3043640}
The numerical solution is denoted by $q_{m}$. 
The discretization error in $q_m$, which is on the order of $10^{-5}$, is much
smaller than the numerical error in the neural network approximation.
Therefore, we neglect the discretization error 
and treat $q_{m}$ as the ``exact'' solution. 
Then the ``exact'' solution for the committor function in $10$D 
is given by $q(x)=q_{m}(x_{1},x_{2})$,
$x\in \Omega =\{x:  (x_1,x_2)\in \tilde\Omega, x\in \mathbb{R}^{10} \}$.
Fig.~\ref{fig2} (top panel) shows the contour plots of the 2D rugged Mueller potential 
and the committor function $q_{m}$ at $k_{B}T=10$ 
(the energy barrier from $A$ to $B$ is about $100$).

Next, we compute the neural network approximation to 
the committor function in the $10$D space using the method proposed 
in section~\ref{method}. The Mueller potential~\eqref{Mueller} 
has two local minima around
$a= (-0.558, 1.441)$ and $b= (0.623, 0.028)$, respectively.
We take the two metastable sets $A$ and $B$ as the cylinders
centered at $(x_1,x_2)=a$ and $(x_1,x_2)=b$ respectively, 
each with radius $r=0.1$. The function $\chi_{A}(x)$ is constructed as
\begin{equation}\label{indicator}
 \chi_{A}(x) = \frac{1}{2}-\frac{1}{2}\tanh[1000 
  (\lvert (x_{1},x_{2})-a\rvert^2 -(r+0.02)^2)],\quad x\in \mathbb{R}^{10}
\end{equation}
and similarly for $\chi_B(x)$. These two functions satisfy \eqref{asum} approximately.

The data are sampled using the two different methods discussed 
in section \ref{importance}.
In the first one, we generate the data at the artificial temperature $k_{B}T'=20$
by solving the Langevin equation~\eqref{lag}
using the Euler-Maruyama scheme with the time step $\Delta t=10^{-5}$.
In the second method, we use metadynamics to generate the data. 
We first bias the coordinates $x_{1}$ and $x_{2}$ 
(i.e. $S_1(x)=x_1$, $S_2(x)=x_2$ in Eq. \eqref{meta_po})
by adding $2000$ Gaussian functions with height $w=5$ and 
width $\sigma_{1}=\sigma_{2}=0.05$ into the potential, one for every 500 time steps. 
Then a set of data are sampled by simulating the Langevin dynamics 
on the modified potential with the time step $\Delta t=10^{-5}$.

In both methods, we take one sample for every $100$ time steps, 
and only keep those data points with coordinates $x\in \Omega\setminus (A\cup B)$.
Of these data, $70\%$ serves as the training dataset and the other $30\%$ serves
as the validation dataset. The neural network used in this example is fully connected, 
and the hyperbolic tangent (tanh) function is used as the activation function 
in the hidden layers. We 
use the package TensorFlow \cite{tensorflow2015-whitepaper}
with Adam optimizer \cite{kingma2014adam} to train the network 
 by minimizing \eqref{newVP2} and \eqref{metaVP} respectively
at the physical temperature $k_B T =10$.



To quantitatively evaluate our results, we compare neural network approximation 
$q_\theta$ with the FEM result $q$. For rare events considered here, 
we are interested in the iso-surface of the committor function where $q\approx 1/2$,
 in particular, the region near the minima of $V$ on this surface. 
This region represents the bottleneck to the transition and 
is known as the transition state region. The committor function in this region 
is also the most difficult to compute, as data here is typically scarce. 
Therefore, we shall focus our evaluation of the numerical results 
in this region. To this end, we carry out the constrained sampling 
in the transition state region by simulating the dynamics
\begin{equation}\label{lag2}
\dot{x}=-\nabla \left( V(x)+V_{q}(x)\right) +\sqrt{2\beta^{-1}}\eta,
\end{equation}
where the additional potential 
$V_{q}(x) =\frac{1}{2}\kappa\left(q_\theta(x)-\frac{1}{2}\right)^2$
with $\kappa=3\times 10^4$ is to constrain the system on the $1/2$-isocommittor surface
$\Gamma_{1/2}=\{x\in\Omega:q_\theta (x)=1/2\}$.
After equilibration, we sample $100$ points: 
$\left\{x^{(k)}: k=1,\dots, 100\right\}$. These points, projected 
on the $(x_1, x_2)$ plane, are shown in Fig.~\ref{fig2} (middle and lower panels). 
The middle panels shows the result obtained using artificial temperature,
and the lower panel shows the result obtained using data sampled from metadynamics. 
The region where these points are clustered is the transition state region.
Also shown in the two panels are comparisons of the $1/2$-isosurface 
of $q_\theta$ with that of $q$ obtained from the FEM calculation. 
Evidently these two agree well in the transition state region.
Certain discrepancies occur away from the transition state region, 
due to the lack of training data there; nevertheless, those regions are 
irrelevant to the transition events.

\begin{table}[t!]
\caption{Comparison of the neural network approximation $q_\theta$
 and the FEM solution $q$ for the extended Mueller potential. 
 The errors are computed on $100$ transition states sampled from 
  the dynamics~\eqref{lag2}. The statistics of the error (mean $\pm$ deviation) 
 is based on $10$ independent runs. 
The network denoted by $10\text{-}20\text{-}20\text{-}1$ contains $2$ hidden layers 
with $20$ nodes on each hidden layer; similarly for other networks.
}
	\label{tab1}
	\begin{tabular}{ c|cccc }
	\hline\hline
	sampling  & \multirow{2}{*}{  data size} & \multirow{2}{*}{network} & \multirow{2}{*}{RMSE} & \multirow{2}{*}{MAE}  \\
	method & & & &\\
	\hline
	\multirow{9}{*}{\shortstack[l]{artificial\\temperature$\ $}} & \multirow{3}{*}{$10^5$} &$10\text{-}20\text{-}1$ &  $0.0607\pm 0.0201$ & $0.0488 \pm 0.0176$ \\
	 & & $10\text{-}20\text{-}20\text{-}1$ & $0.0636\pm 0.0288$ & $0.0519 \pm 0.0240$\\
	& &$\ \ 10\text{-}20\text{-}20\text{-}20\text{-}1\ \ $ & $\ \ 0.0520 \pm 0.0157\ \ $ & $\ \ 0.0425 \pm 0.0135\ \ $ \\
	\cline{2-5}
	& \multirow{3}{*}{$2\times 10^5$} &$10\text{-}20\text{-}1$ &  $0.0456\pm 0.0070$ & 	$0.0368 \pm 0.0058$ \\
	& &$10\text{-}20\text{-}20\text{-}1$ & $0.0427\pm 0.0079$ 	& $0.0345 \pm 0.0065$\\
	& &$10\text{-}20\text{-}20\text{-}20\text{-}1$ & $0.0416 \pm 0.0111$ & $0.0338 \pm 0.0099$ \\
	\cline{2-5}
	& \multirow{3}{*}{$4\times 10^5$} &$10\text{-}20\text{-}1$ &  $0.0261\pm 0.0062$ & 	$0.0213 \pm 0.0054$ \\
	& &$10\text{-}20\text{-}20\text{-}1$ & $0.0353\pm 0.0076$ 	& $0.0297 \pm 0.0074$\\
	& &$10\text{-}20\text{-}20\text{-}20\text{-}1$ & $0.0309 \pm 0.0100$ & $0.0256 \pm 0.0095$ \\
	\hline
	\hline
	\multirow{3}{*}{\shortstack[l]{meta-\\dynamics}} & $10^4$ &$10\text{-}20\text{-}1$ &  $0.0737 \pm 0.0169$ & $0.0572 \pm 0.0135$ \\
	\cline{2-5}
	& $2\times10^4$ &$10\text{-}20\text{-}1$ &  $0.0532 \pm 0.0094$ & $0.0441 \pm 0.0081$ \\
	\cline{2-5}
	& $4\times10^4$ &$10\text{-}20\text{-}1$ &  $0.0345 \pm 0.0087$ & $0.0285 \pm 0.0081$ \\
	\hline\hline
	\end{tabular}
\end{table}

We also compute the root-mean-square error (RMSE)
and the mean-absolute error (MAE) between $q_\theta$ 
and the FEM solution $q$ at the sampled transition states $x^{(k)}, k=1,..., 100$,
\begin{align*}
\text{RMSE} & = \sqrt{\frac{1}{N_s}\sum_{k=1}^{N_s}
\left(q_\theta(x^{(k)})-q(x^{(k)}) \right)^2 }, \\
\text{MAE} & = \frac{1}{N_s}\sum_{k=1}^{N_s}
\left| q_\theta(x^{(k)})-q(x^{(k)}) \right|,
\end{align*}
where $N_s=100$. The results for different dataset and difference choice of
network structure are reported in Table~\ref{tab1}. 
Each error shown in the table is computed from $10$ independent runs; 
each run includes sampling the data, training the neural network, 
and sampling the transition states.
We observe that the numerical results are insensitive to 
the number of hidden layers, but the accuracy improves for larger set of training data.
We also observe that the sampling method based on metadynamics is more efficient than the method of artificial temperature. 
To achieve roughly the same accuracy, the latter requires one order 
of more data as compared to the method based on metadynamics. 
This is due to the fact that a larger portion of training data from 
metadynamics are concentrated in the transition state region.

\subsection{Alanine Dipeptide}

In this example, we study the isomerization process of the alanine
dipeptide in vacuum at $T=\SI{300}{\kelvin}$. The isomerization of alanine dipeptide has been the subject of several theoretical and computational studies, \cite{apostolakis1999calculation,Bolhuis2000,ma2005automatic,ren2005transition}
therefore it serves as a good benchmark problem for the proposed method.

\begin{figure}[t!]
	\centering
		\includegraphics[width=.6\linewidth]{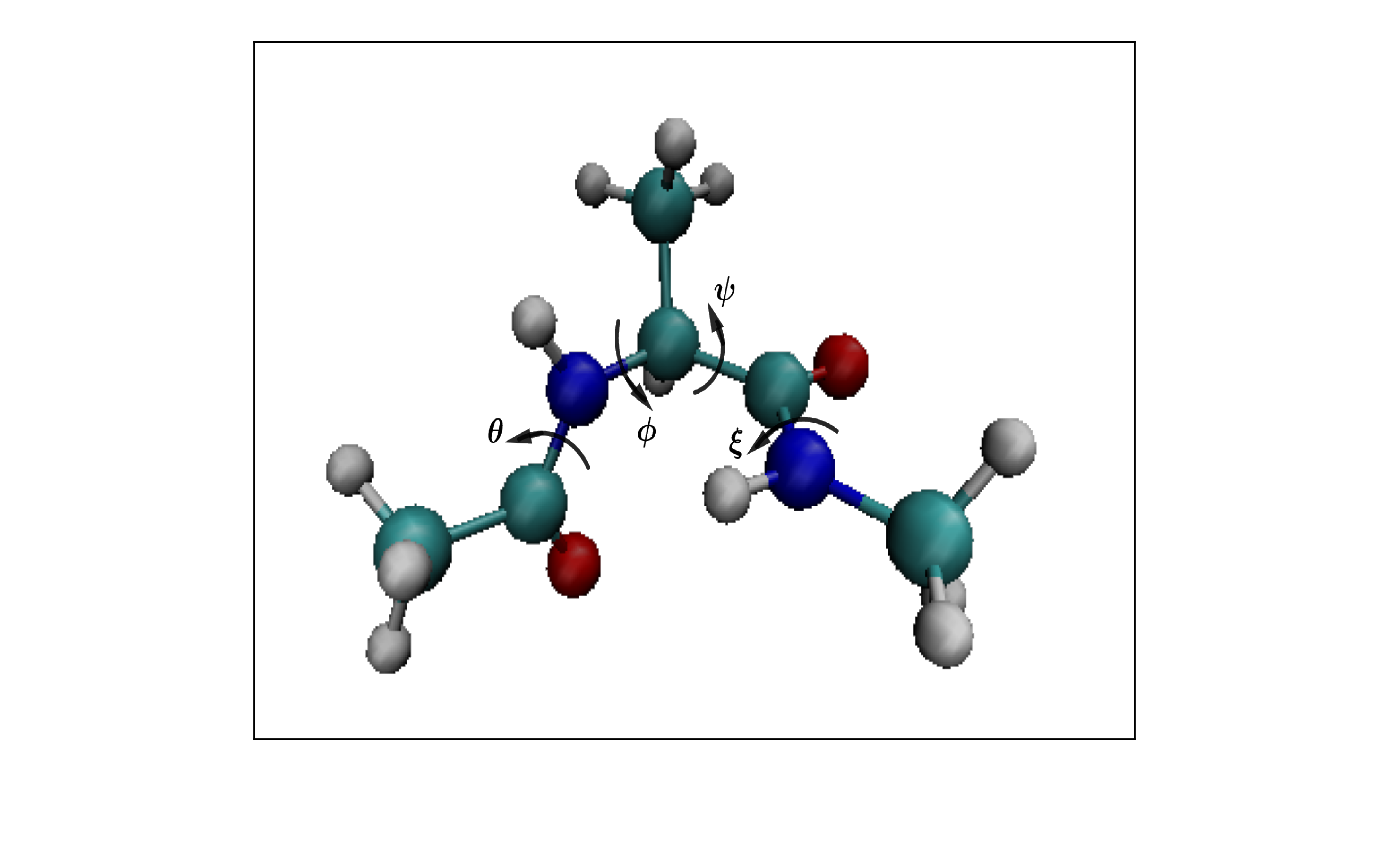}
		\includegraphics[width=.6\linewidth]{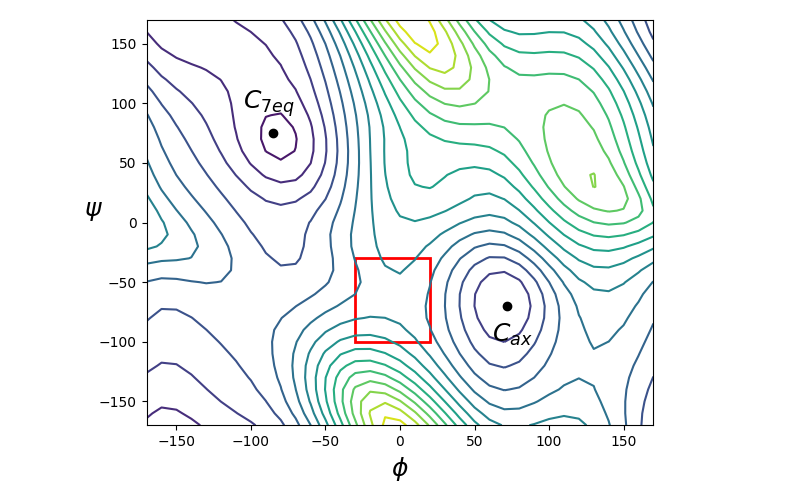}
\caption{{\it Upper panel:} Schematic representation of the alanine dipeptide
($\text{CH}_{3}$$\text{-}$$\text{CONH}$$\text{-}$$\text{CHCH}_{3}$$\text{-}$$\text{CONH}$$\text{-}$$\text{CH}_{3}$).
{\it Lower panel:} The adiabatic energy landscape of alanine dipeptide 
obtained by minimizing the potential energy of the molecule with the two torsion angles $\phi$ and $\psi$ fixed. The small rectangle indicates 
the transition state region observed in earlier studies.
}
	\label{fig3}
\end{figure}

The molecule consists of 22 atoms and has a simple chemical structure,
yet it exhibits some of the important features common to biomolecules.
Figure~\ref{fig3} shows the stick and ball representation of the molecule (upper panel) 
and its adiabatic energy landscape on the plane of the two torsion 
angles $\phi$ and $\psi$ (lower panel). 
The molecule has two metastable conformers $C_{7eq}$ and $C_{ax}$ located around $(-\ang{85},\ang{75})$ and $(\ang{72},-\ang{75})$, respectively.
Accordingly, the metastable sets $A$ and $B$ are chosen as
\begin{equation}
\begin{aligned}
A &= \left\{x: |(\phi(x),\psi(x))-C_{7eq}|<\ang{10}\right\},\\
B &= \left\{x: |(\phi(x),\psi(x))-C_{ax}|< \ang{10}\right\}.
\end{aligned}
\end{equation}
Our goal is to compute the committor function for transitions between $A$ to $B$ 
and sample the transition states at the temperature $T=\SI{300}{\kelvin}$.

We use the two sampling methods discussed in section \ref{importance} 
to generate the training data. In the method of artificial temperature, 
we raise the temperature to $T'=\SI{800}{\kelvin}$ and use 
the package NAMD \cite{phillips2005scalable} 
to simulate the Langevin dynamics with the time step $\Delta t=\SI{0.5}{\fs}$.
In the method of metadynamics, we bias the two torsion angles $(\phi,\psi)$ 
(i.e. $S_1(x)=\phi$, $S_2(x)=\psi$ in Eq. \eqref{meta_po}). 
We add one Gaussian function with height $w=\SI{0.1}{\kcal\per\mol}$ 
and width $\sigma_{1}=\sigma_{2}=\ang{8}$ for every 500 time steps.
In total, $10^4$ Gaussian functions are added to fill the potential wells 
containing the metastable sets $A$ and $B$. 
To save the computational cost,  \cite{fiorin2013using} the biased potential and its gradient are accumulated and computed on a uniform grid with mesh size $\ang{2}$ for $\phi$ and $\psi$. 

\begin{figure}[t!]
\centering
\includegraphics[width=.6\linewidth]{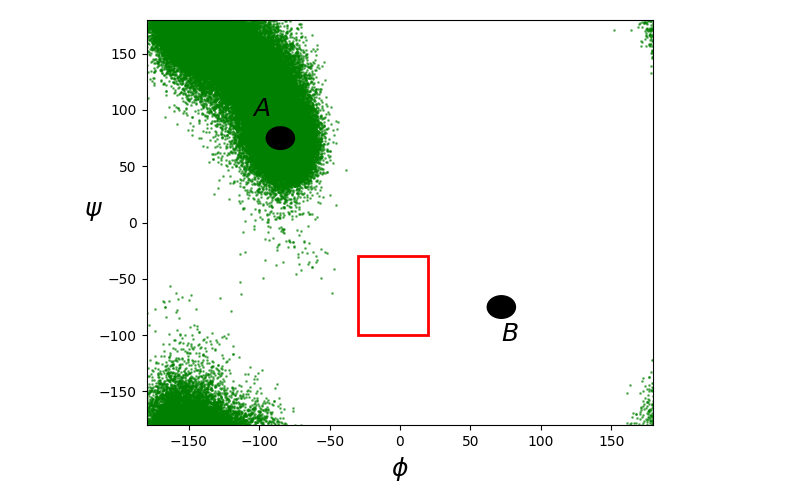}
\includegraphics[width=.6\linewidth]{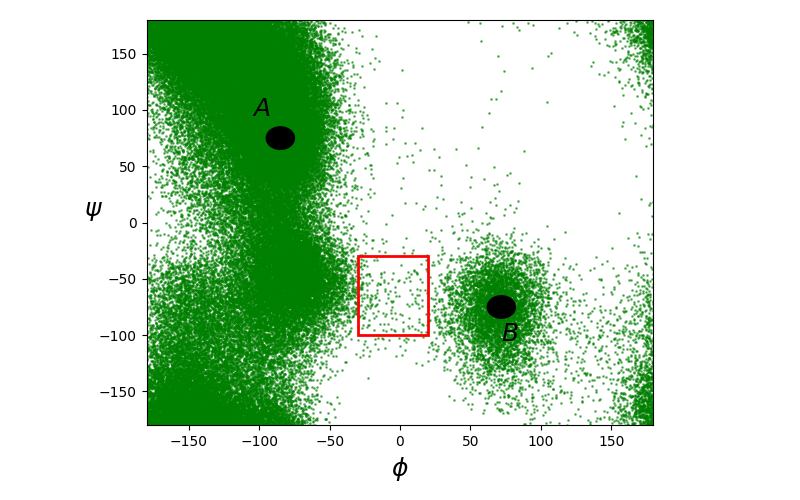}
\includegraphics[width=.6\linewidth]{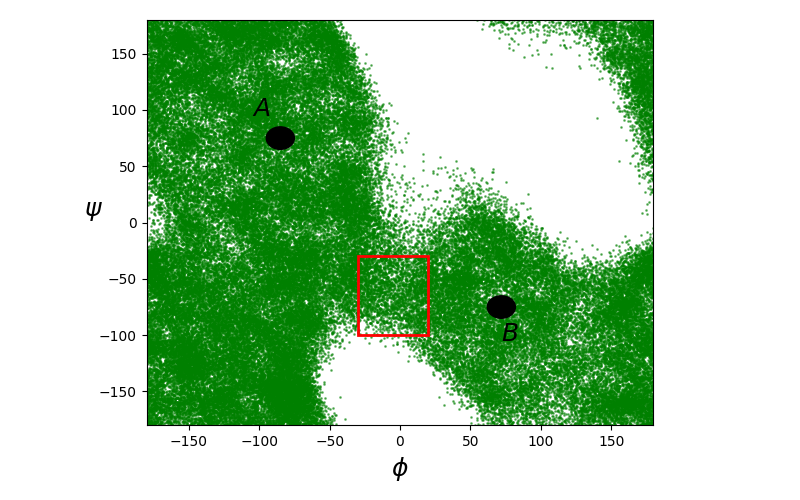}
\caption{Distribution of $10^5$ data points sampled from 
the Langevin dynamics of alanine dipeptide at the temperature $T=\SI{300}{\kelvin}$ 
(upper panel), $T'=\SI{800}{\kelvin}$ (middle panel), 
and from metadynamics (lower panel). The unit of the coordinates is degree.}
	\label{fig4}
\end{figure}

In Fig.~\ref{fig4}, we show the data sampled at the temperature $T'=\SI{800}{\kelvin}$
(middle panel) and from metadynamics on the modified potential 
at the physical temperature $T=\SI{300}{\kelvin}$ (lower panel), respectively. 
These data are sampled from $10^7$ steps of Langevin dynamics, one for every 100 steps. 
We observe that the data sampled from metadynamics has better quality in the sense
that a larger portion of the data are in the transition state region.
As a comparison, we also run the dynamics for the same number of time steps
on the original potential at the temperature $T=\SI{300}{\kelvin}$,
and plot the sampled data in the figure (upper panel). 
It is seen that at the temperature $T=\SI{300}{\kelvin}$, 
no transition from $C_{7eq}$ to $C_{ax}$ is observed within the
simulation time; consequently, no training data is collected 
in the transition state region, the region of our interest.
This demonstrates the benefit of our proposed sampling methods.

In the design of the network structure, we use feature engineering 
via collective variables. As conformational changes of bio-molecules 
can often be well described by a few collective variables such as torsion angles, 
we design the first layer of our network to
extract a collection of such torsion angles from the molecule,
whose sines and cosines are then fed into the subsequent hidden layers 
as extracted features. Note that these include the four torsion angles shown 
in Fig. \ref{fig3} that were identified as being adequate in describing the transition
in earlier work (e.g. Ref.~\onlinecite{maragliano2006string}).
But we do not \textit{a priori} assume that we know this precise information, 
and a redundant description is supplied.

We train the network at $T=\SI{300}{\kelvin}$ using $2\times 10^6$ data points sampled at $T'=\SI{800}{\kelvin}$ and $10^5$ data points sampled from metadynamics, respectively.
All the data are sampled outside the sets $A$ and $B$. 
Of these data, $70\%$ serves as the training data and $30\%$ are used for validation.
With these data, we use the package TensorFlow \cite{tensorflow2015-whitepaper}
with Adam optimizer \cite{kingma2014adam} to minimize the objective functions in
 \eqref{newVP2} and \eqref{metaVP}, respectively.
 The computation is terminated when the validation error no longer decreases.



\begin{figure}[t!]
	\begin{subfigure}[t]{.5\textwidth}\centering
		\includegraphics[width=\linewidth]{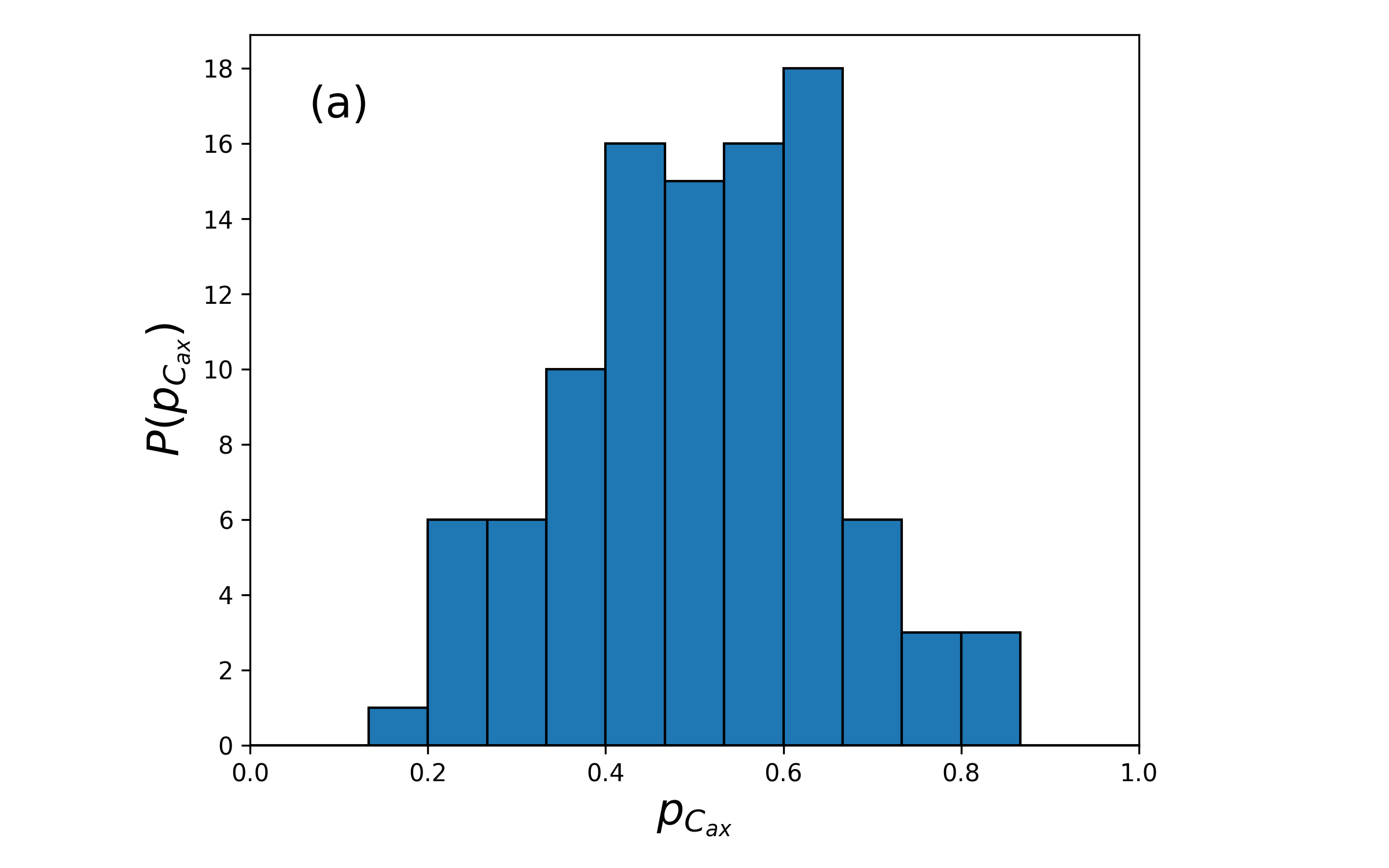}
	\end{subfigure}%
	\begin{subfigure}[t]{.5\textwidth}\centering
		\includegraphics[width=\linewidth]{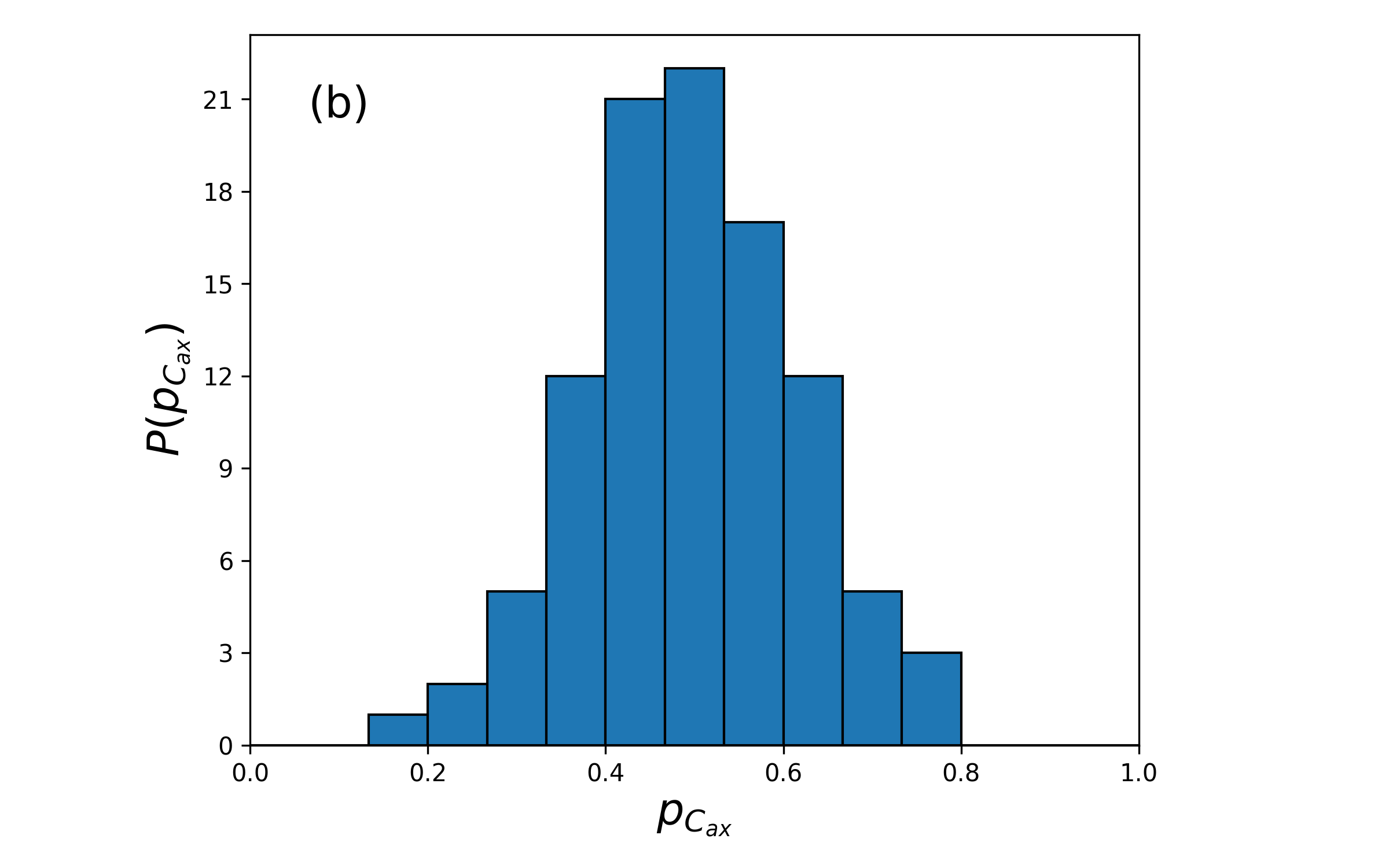}
	\end{subfigure}
	\begin{subfigure}[t]{.5\textwidth}\centering
		\includegraphics[width=\linewidth]{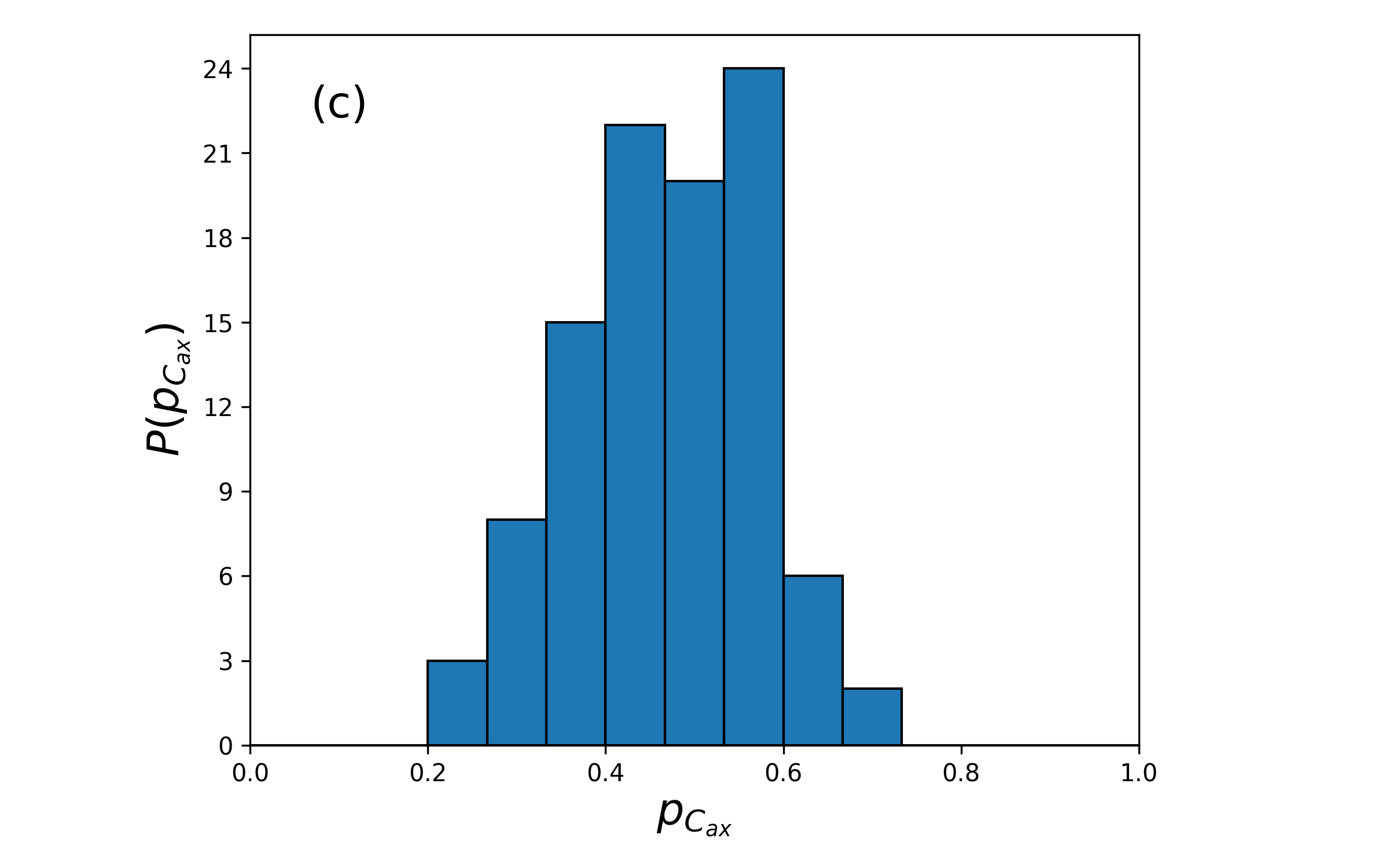}
	\end{subfigure}%
	\begin{subfigure}[t]{.5\textwidth}\centering
		\includegraphics[width=\linewidth]{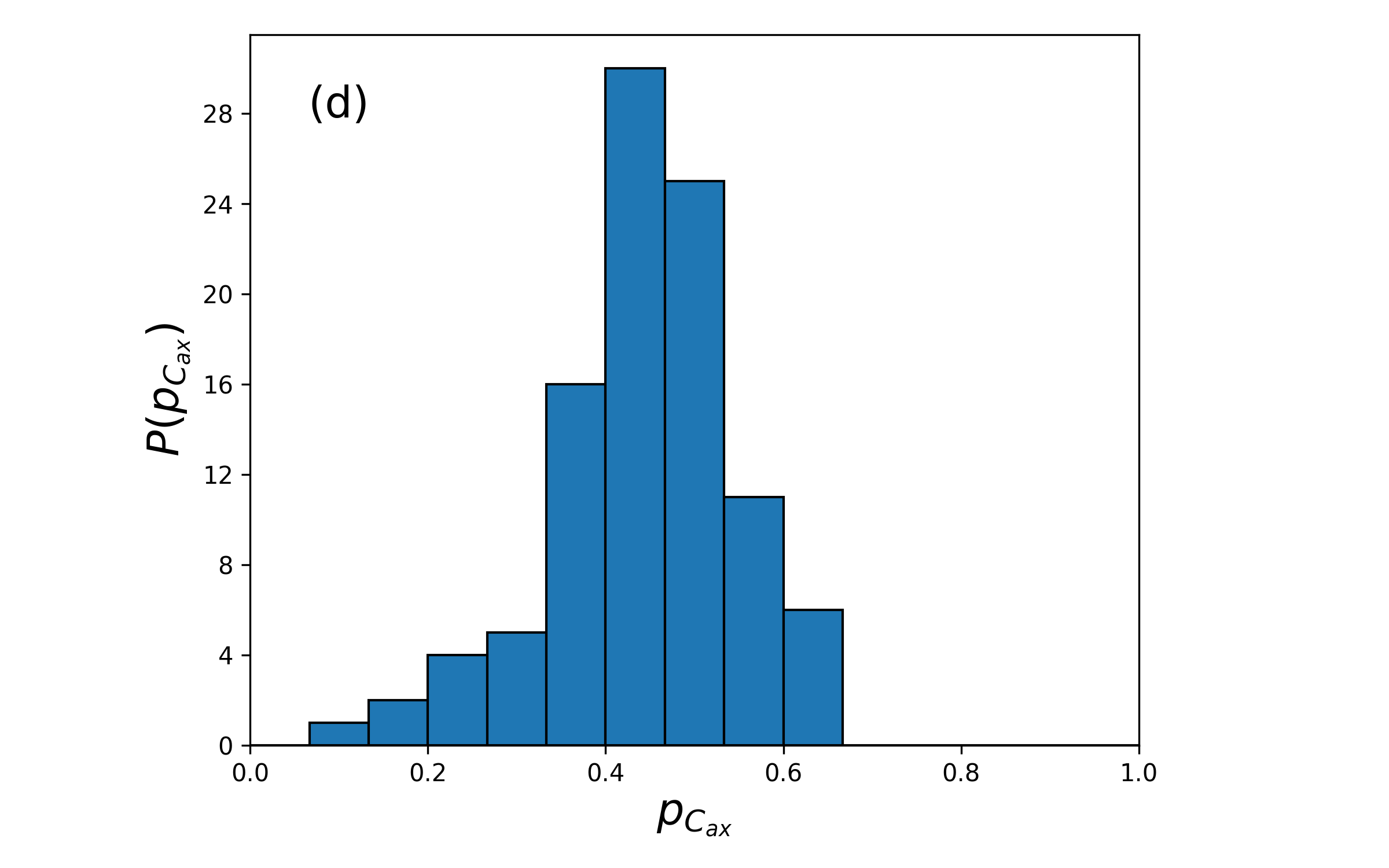}
	\end{subfigure}
		\begin{subfigure}[t]{.5\textwidth}\centering
		\includegraphics[width=\linewidth]{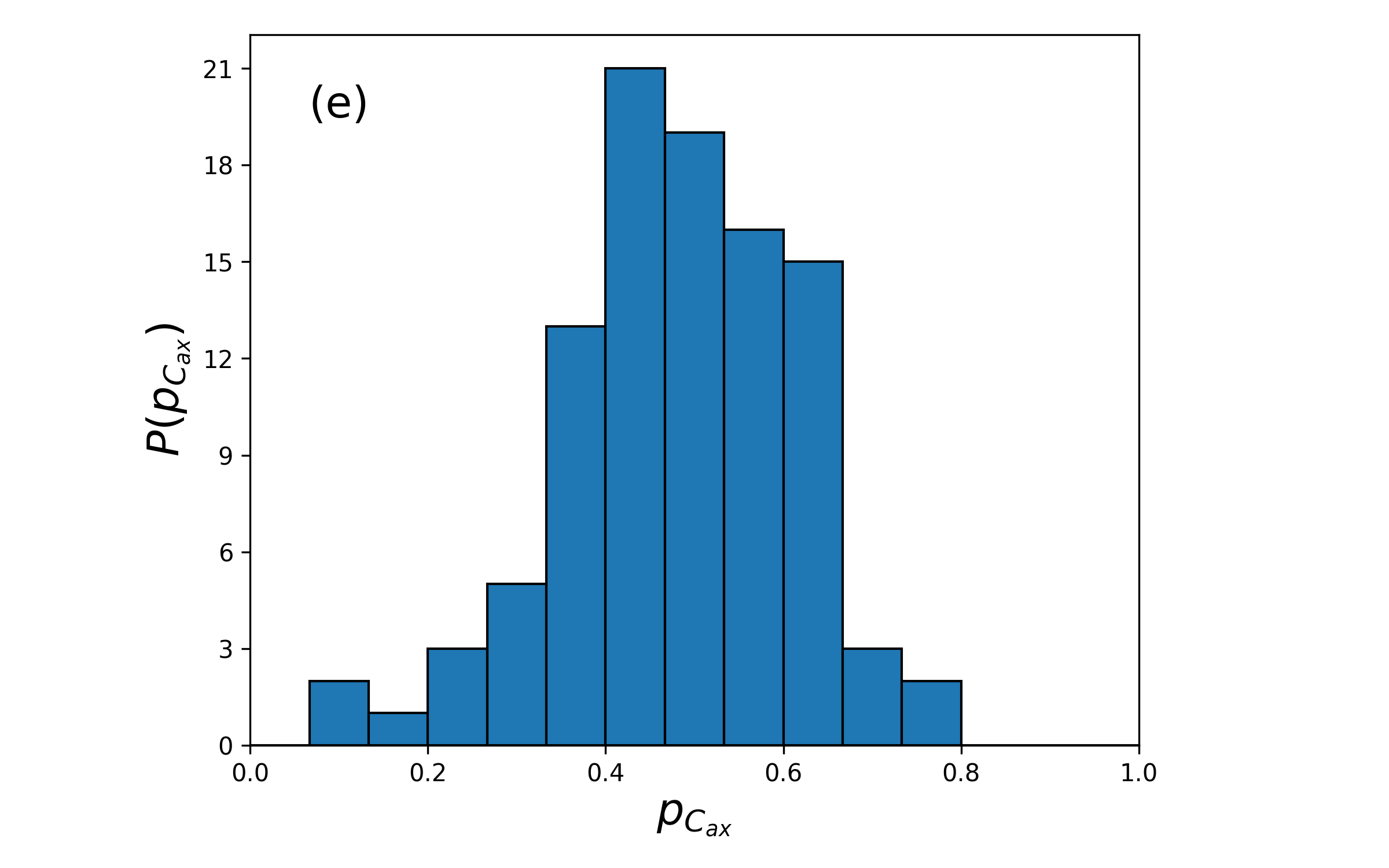}
	\end{subfigure}%
	\begin{subfigure}[t]{.5\textwidth}\centering
		\includegraphics[width=\linewidth]{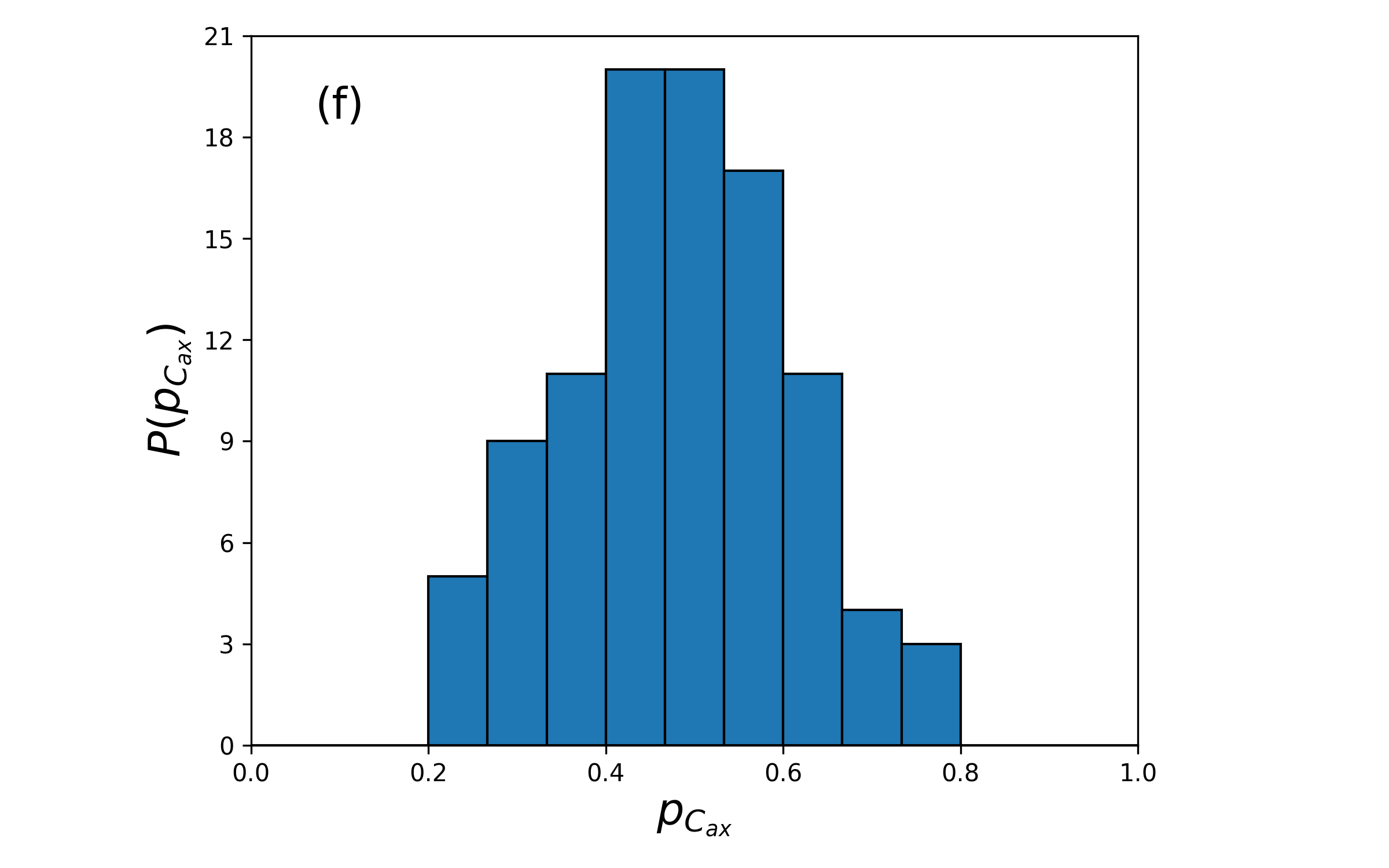}
	\end{subfigure} 
\caption{Distribution of the committor values for the $100$ states sampled 
on the $1/2$-isosurface of $q_\theta$. The committor values are computed directly
using the Langevin dynamics.  
{\it (a)-(b)}: Two independent runs using the sampling method of artificial temperature at $T'=\SI{800}{\kelvin}$ and using a $66\text{-}18\text{-}30\text{-}1$ network 
with $9$ torsion angles for the feature layer.
{\it (c)-(d)}: Two independent runs using the sampling method of metadynamics and using a $66\text{-}18\text{-}30\text{-}1$ network with $9$ torsion angles for the feature layer.
{\it (e)-(f)}: Two independent runs using the sampling method of metadynamics and using a $66\text{-}82\text{-}100\text{-}1$ network with $41$ torsion angles for the feature layer.}
	\label{fig5}
\end{figure}

For this high-dimensional problem, we cannot afford to compute the committor function using the FEM method as we did in the first example.
In order to check the accuracy of the numerical results, 
in particular, whether the $1/2$-isocommittor surface really 
locates the transition states, we carry out the constrained Langevin dynamics 
simulation on the $1/2$-isocommittor surface at $T=\SI{300}{\kelvin}$.
Following the dynamics, we collect $100$ states.
The committor values of these states are then computed directly using Langevin dynamics.
Specifically, we generate $200$ trajectories initiating from each of these states 
with random initial velocities and estimate the probability (i.e. the committor value) 
of the system first reaching $B$ rather than $A$ (c.f.~\eqref{qdef}).
If adequate accuracy is achieved, we would expect the computed probabilities to cluster around $1/2$. We carry out two independent runs, including sampling data,
training the network and sampling the transition states, for each of the
two sampling methods. Fig.~\ref{fig5} shows the distribution 
of the committor values at the sampled transition states. 
Results in (a)-(d) and (e)-(f) are obtained using 9 and 41 torsion angles 
in the feature layer of the network, respectively. In all the results, 
 the committor values cluster around $q=1/2$, indicating the transition states
are correctly identified. 
We also tested neural networks with fewer nodes in the hidden layer and obtained similar results.



\section{Conclusion}
\label{conclusion}

In this paper, we introduced a method for computing the committor function at low temperatures. The committor function characterizes rare transition events between metastable states. The main idea of the method is to combine deep learning with efficient sampling methods in order to overcome the curse of dimensionality associated with realistic systems 
and the scarcity of transition data at low temperatures. 
We also incorporated collective variables into the network structure 
as a form of crude feature selection to improve the efficiency of learning.

We considered two sampling methods: the method of artificial temperature and the method based on metadynamics. In the numerical examples considered in this work, the method of metadynamics outperformed the sampling method at artificial temperature.
The former produced training data of better quality in the sense that a larger portion of the data lie in the transition state region. As a result, less amount of data were needed to achieve the same accuracy as in the sampling method using higher temperatures. Nevertheless, metadynamics requires a suitable choice 
of collective variables to bias, while sampling at higher temperature 
does not need this information.

Our method is demonstrated to be effective on a relatively simple example involving the rugged Mueller potential, as well as a more complex benchmark example of the alanine dipeptide molecule. This provides an alternative approach to study complex systems with rough energy landscapes. We intend to apply the method to more complex systems in the future work.

\begin{acknowledgments}
	The work of Ren was supported in part by Singapore MOE AcRF grants 
	R-146-000-232-112 (Tier 2) and R-146-000-267-114 (Tier 1), 
	and the NSFC grant (No. 11871365).
\end{acknowledgments}

\section*{reference}
\bibliography{v2}

\end{document}